\documentclass[twocolumn,showpacs,amsmath,amssymb,prl]{revtex4}
\usepackage{graphicx}
\usepackage{bm}
\usepackage{psfig}
\begin{document}

\title{High-quality quantum point contact in two-dimensional GaAs (311)A hole system}

\author{J.~Shabani$^{1}$}

\author{J.~R.~Petta$^{2}$}

\author{M.~Shayegan$^{1}$}

\affiliation{$^{1}$Department of Electrical Engineering, Princeton
University, Princeton, NJ 08544, USA
\\
$^{2}$Department of Physics, Princeton University, Princeton, NJ
08854, USA}

\date{\today}

\begin{abstract}

We studied ballistic transport across a quantum point contact
(QPC) defined in a high-quality, GaAs (311)A two-dimensional (2D)
hole system using shallow etching and top-gating. The QPC
conductance exhibits up to 11 quantized plateaus. The ballistic
one-dimensional subbands are tuned by changing the lateral
confinement and the Fermi energy of the holes in the QPC. We
demonstrate that the positions of the plateaus (in gate-voltage),
the source-drain data, and the negative magneto-resistance data
can be understood in a simple model that takes into account the
variation, with gate bias, of the hole density and the width of
the QPC conducting channel.


\end{abstract}
\pacs{Valid PACS appear here}

\maketitle

 The quantized plateaus observed in the conductance of a narrow
constriction, the so-called quantum point contact (QPC), are the
hallmarks of ballistic transport in a one-dimensional (1D) system
\cite{vanWees88,Wharam88}. The QPC structure is also an essential
building block for many other types of structures in mesoscopic
physics, such as rings and quantum dots. Since the first
observation of conductance quantization in QPCs based on
GaAs/AlGaAs electrons \cite{vanWees88,Wharam88}, the quantization
has been reported in electron systems in several other materials
such as SiGe \cite{SiGe1,SiGe2}, GaN \cite{GaN}, and AlAs
\cite{AlAs}. For a long time, there has also been interest in
ballistic 1D transport in GaAs $hole$ systems
\cite{Zailer94,Daneshvar97,Rokhinson02,Danneau06,Koduv08}, partly
because of the stronger spin-orbit interaction and the resulting
spin-splitting of the energy bands at finite wavevectors
\cite{Winkler03}. The realization of high-quality and stable QPCs
in hole systems, however, has been challenging. Here we report on
an exceptionally high-quality GaAs hole QPC device in which we
observe up to 11 conductance steps and the ``0.7" structure. We
present a simple model that explains the gate-voltage dependance
of the plateaus' positions as well as the source drain bias data
and the magneto-resistance of the QPC. Spacings between the
quantized energy levels in this geometry are about 5 times larger
than in previous reports of QPCs in other GaAs hole systems.
\begin{figure}[h!]
\centering
\includegraphics[scale=0.58]{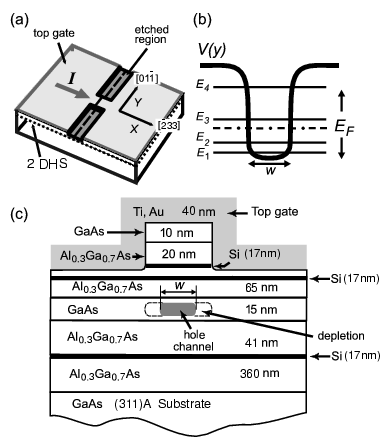}
\caption{(a) Device schematic of the shallow-etched QPC. The
crystal orientation is also shown. (b) Confinement potential along
the dashed line in (a). The Fermi energy ($E_{F}$) and the width
of the QPC channel ($w$) vary with the gate voltage bias. (c)
Device cross section along the dashed line in (a).}
\end{figure}

Our QPC device is based on a constriction made in a high-mobility,
GaAs two-dimensional hole system (2DHS). The 2DHS is realized in a
modulation-doped structure, grown on a GaAs (311)A substrate by
molecular beam epitaxy. The 2DHS is confined to a GaAs quantum
well flanked by un-doped (spacer) and Si-doped layers of
Al$_{0.3}$Ga$_{0.7}$As. The top Al$_{0.3}$Ga$_{0.7}$As/GaAs
interface is 112 nm below the surface. The layer sequence of the
sample is schematically shown in Fig. 1(c) \cite{footnote1}. The
2DHS has a typical low-temperature mobility of $5 \times
10^5$~cm$^2$/Vs along the $[01\bar{1}]$ direction and $7 \times
10^5$~cm$^2$/Vs along $[\bar{2}33]$ at a 2D hole density $p_{2D}$
of $3.8 \times 10^{11}$~cm$^{-2}$. The QPC is defined with a
lithographic width of 300 nm and length of 400 nm. Figure 1(a)
shows the sample geometry. The sample is etched 40 nm deep, using
an electron cyclotron resonance plasma etcher with ZEP 520A resist
as an etch mask, to fabricate a QPC with a strong confinement
potential (Fig. 1b). A Ti/Au gate is then deposited on the top of
the sample, covering the QPC channel as well as the adjacent 2DHS
reservoirs. Applying a voltage bias, $V_{G}$, to this gate changes
the hole density in both the QPC channel and the reservoirs. We
made measurements in a dilution refrigerator using the standard
low-frequency AC lock-in techniques with an excitation of 10
$\mu$V at 17 Hz.

Figure~2 shows the conductance, $G$, of the QPC device, measured
at a temperature of $T$ = 70 mK as a function of $V_{G}$. The QPC
conductance exhibits 11 clear steps, most of which are
well-quantized in multiple integers of $2e^{2}/h$. Note that in
our structure, increasing $V_{G}$ leads to a narrowing of the
constriction and also a lowering of the hole density in the 2D
reservoirs and the QPC channel. As a result, the 1D hole subbands
are depopulated one-by-one as $V_{G}$ is increased, and
conductance plateaus are observed at $(2e^{2}/h)N$, where $N$ is
an integer denoting the number of occupied 1D subbands. The
plateaus in our sample are quantized at the expected values for $0
\leq N \leq 11$, except for $N$ = 7, 8, and 9. We do not know the
origin of this anomaly, but we note that qualitatively similar
anomalies have been observed in previous hole QPCs
\cite{Rokhinson02,Koduv08}. In addition to the last non-zero
plateau at $2e^{2}/h$, the device also shows the ``0.7" structure
which is more pronounced at higher temperatures as shown in the
inset of Fig.~2.
\begin{figure}
\centering
\includegraphics[scale=0.6]{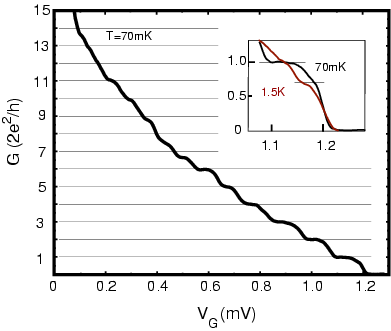}
\caption{QPC conductance ($G$) vs. gate bias ($V_{G}$), showing 11
conductance plateaus, most of which are quantized in integer
multiples of $2e^{2}/h$. The inset provides an expanded view of
the region between the first two plateaus, illustrating the
observation of the ``0.7" structure in this QPC.}
\end{figure}

The separation between the 1D subband energies in a QPC channel
can be extracted from measurements of the nonlinear conductance as
a function of the source-drain bias, $V_{SD}$, applied across the
QPC. In Fig.~3 we plot the differential transconductance of our
device as a function of $V_{G}$ and the DC $V_{SD}$
\cite{footnote2}. When $V_{SD}$ is equal to the energy difference
between the $N$ and $N+1$ subbands, an additional quantized
conductance plateau with a value equal to an (2$N$+1)/2 times
$2e^{2}/h$ appears. The subband energy spacing $\Delta E_{N,N+1}$
can then be deduced from the $V_{SD}$ values at which these
half-odd-integer plateaus are centered in Fig.~3. We find the
energy spacings $\Delta E_{1,2}=0.7$ meV, $\Delta E_{2,3}=0.6$
meV, $\Delta E_{3,4}=0.55$ meV, and $\Delta E_{4,5}=0.5$ meV in
our system. These are about 5 to 7 times larger than $\Delta
E_{N,N+1}$ reported for other hole QPCs \cite{Danneau06,Koduv08},
likely because of the strong confinement potential in our device
due to etching. Compared to the energy spacings in GaAs $electron$
samples with similar, side-etched, QPC structures of Kristensen
$et$~$al.$ \cite{Kristensen00}, on the other hand, the spacings in
our QPC are about 4 times smaller, qualitatively consistent with
the much larger effective mass of holes compared to the electron
effective mass.

\begin{figure}
\centering
\includegraphics[scale=0.65]{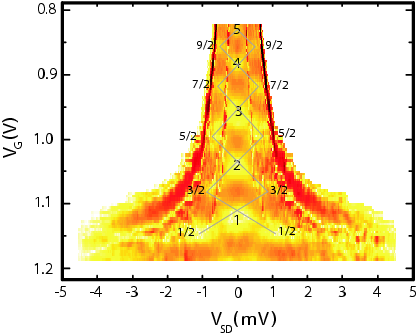}
\caption{(Color online) Color map of the differential
transconductance as a function of gate bias,~$V_{G}$, and applied
DC source-drain bias,~$V_{SD}$. The data have been corrected for
the voltage drop across series resistance. the plateaus appear as
the light regions (yellow) while darker regions (red) represent
the rising conductance between the plateaus.}
\end{figure}

Next, we describe a simple model for the potential landscape
around the QPC, and compare its predictions to the experimental
data. The model is similar to the one used by Gunawan $et$ $al.$
\cite{AlAs} to explain data for an AlAs electron QPC with a
similar geometry, namely, a QPC which is side-etched and whose
density is varied by biasing a top gate. Note that in our geometry
(Fig.~1), as we decrease $V_{G}$, $E_{F}$ increases and crosses
the quantized energy levels in the QPC, leading to quantized
conductance steps. We can deduce $E_{F}$ from the density
($p_{2D}$) of the 2DHS reservoirs that surround the QPC; $p_{2D}$,
in turn, can be determined from measurements of the Shubnikov-de
Haas oscillations and the Hall resistance in the 2DHS reservoirs.
Since the QPC confinement potential is strong and the quantized
energies are small thanks to the large hole effective mass, we can
estimate the 1D subband energy levels using an infinite square
well model: $E_{N}=N^{2}\pi^{2}\hbar^{2}/2m^{*}w^{2}$, where
$m^{*}$ is the hole effective mass and $w$ is the width of the
(conducting) QPC channel. Assuming $m^{*}=~0.2m_{e}$, where
$m_{e}$ is the electron mass, we can then deduce the value of $w$
at the consecutive crossings of $E_{F}$ and $E_{N}$. The width
derived from this model is plotted in Fig.~4 as open circles. Now,
it is also possible to estimate the channel width from the
magneto-resistance of the QPC. When the QPC is subjected to a
small perpendicular magnetic field, its resistance exhibits a
strong negative magneto-resistance which ends with a pronounced
kink at a characteristic field $B_{K}$, marking the field at which
the classical cyclotron diameter equals the channel width
\cite{AlAs,kink,kink2}. Using $B_{K}$ one can then estimate the
width of the QPC: $w=2\hbar k_{F}/eB_{K}$, where $k_{F}$ is the
Fermi wavevector. The filled diamonds in Fig.~4 show the width
deduced from the kink positions in the magneto-resistance data for
our sample; the error bars represent the uncertainty in
determining the kink positions. Note that we have ignored the
complications in hole bands arising from the spin-orbit
interaction, i.e., we assumed that the holes in the QPC are spin
degenerate and have a circular Fermi contour. Despite these
simplifications, the channel widths deduced from two independent
sets of data, namely the $V_{G}$ positions of the conductance
plateaus and the field positions of the magneto-resistance kinks,
are in good agreement.
\begin{figure}
\centering
\includegraphics[scale=0.5]{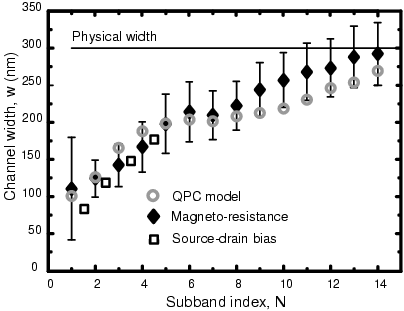}
\caption{The QPC channel width ($w$), deduced from the kink in the
low-field magneto-resistance data (diamonds), from the positions
of the conductance plateaus (open circles), and from the
source-drain measurements (open squares).}
\end{figure}
The source-drain bias data provide a third, independent measure of
$w$:  The energy level spacing in the QPC is
$(2N+1)\pi^{2}\hbar^{2}/(2m^{*}w^{2})$. Open squares in Fig.~4
represent $w$ deduced from source-drain data and are plotted vs.
the average of each two consecutive $N$. These width are also in
agreement with $w$ determined from the other two techniques. It is
also noteworthy that the deduced $w$ are smaller than the physical
(lithographical) width of the QPC and decrease with decreasing
subband index; this behavior is reasonable, as we expect that the
depletion areas near the QPC channel grow as the hole density is
lowered \cite{AlAs}.


 Our work was supported by the NSF MRSEC and DOE. We thank B. H. McGuyer for help with the
measurements.



\begin{thebibliography}{99}
\bibitem{footnote1}
The 17 nm-thick Si-doping layers on the sides of the 15 nm GaAs
quantum well have a Si densitiy of $6.5 \times 10^{17}$ cm$^{-3}$
and the top doping layer has a Si density of $1.75 \times 10^{18}$
cm$^{-3}$.
\bibitem{footnote2}
The data in Fig.~3 have been corrected for the voltage drop across
the series resistance of the cryostat filtered wires.
\end{thebibliography}
\end{document}